# Visualizing the microscopic coexistence of spin density wave and superconductivity in underdoped $NaFe_{1-x}Co_xAs$

Peng Cai[1], Xiaodong Zhou[1], Wei Ruan[1], Aifeng Wang[2], Xianhui Chen[2], Dung-Hai Lee[3], and Yayu Wang[1,*]

[1]*State Key Laboratory of Low Dimensional Quantum Physics, Department of Physics, Tsinghua University, Beijing 100084, P. R. China*

[2]*Hefei National Laboratory for Physical Science at Microscale and Department of Physics, University of Science and Technology of China, Hefei, Anhui 230026, P.R. China*

[3]*Department of Physics, University of California at Berkeley, Berkeley, CA 94720, USA*

[†] Email: yayuwang@tsinghua.edu.cn

**Although the origin of high temperature superconductivity in the iron pnictides is still under debate, it is widely believed that magnetic interactions or fluctuations play an important role in triggering Cooper pairing[1,2]. Because of the relevance of magnetism to pairing, the question of whether long range spin magnetic order *can* coexist with superconductivity *microscopically* has attracted strong interests[3-9]. The available experimental methods used to answer this question are either bulk probes or local ones without control of probing position, thus the answers range from mutual exclusion to homogeneous coexistence. To definitively answer this question, here we use scanning tunneling microscopy to investigate the local electronic structure of an underdoped $NaFe_{1-x}Co_xAs$ near the spin density wave (SDW) and superconducting (SC) phase boundary. Spatially resolved spectroscopy directly reveal both the SDW and SC gap features at the same atomic location, providing compelling evidence for the microscopic coexistence of the two phases. The strengths of the SDW and SC features are shown to anti correlate with each other, indicating the competition of the two orders. The microscopic coexistence clearly indicates that Cooper pairing occurs when portions of the Fermi surface (FS) are already gapped by the SDW order. The regime $T_C < T < T_{SDW}$ thus show a strong resemblance to the pseudogap phase of the cuprates where growing experimental evidences suggest a FS reconstruction due to certain density wave order[10-12]. In this phase of the pnictides, the residual FS has a favorable topology for magnetically mediated pairing when the ordering moment of the SDW is small.**

The evolution between the SDW and SC phases in the underdoped regime of the iron pnictides has been extensively investigated, but a consistent picture is still lacking. Different experimental probes on various iron pnictide compounds have revealed three categories of behavior. Muon spin relaxation (μSR) and Mössbauer spectroscopy on $LaFeAsO_{1-x}F_x$ suggests that the abrupt transition from SDW to SC is first-order-like without coexistence of the two phases[4]. On the contrary, μSR on $SmFeAsO_{1-x}F_x$ (Ref. 5) and $BaFe_{2-x}Co_xAs_2$ (Ref. 8), together with nuclear magnetic resonance[6] and neutron scattering[7] on similar compounds, reveal the coexistence of the two phases. In an intermediate case, neutron scattering in $CeFeAsO_{1-x}F_x$ reveals a continuous transition, but the SDW and SC regimes only touches at a quantum critical point[3]. Among these scenarios the possible microscopic coexistence of magnetism and superconductivity has attracted wide attention for its possible implication on pairing symmetry and mechanism[13-15].

With its atomic scale structural and spectroscopic imaging capabilities, scanning tunneling microscopy (STM) is an ideal probe for resolving the issue of microscopic phase coexistence in complex materials. In high $T_C$ cuprates, for example, STM has revealed the microscopic coexistence and distinct temperature/doping/spatial evolutions of the pseudogap and SC phases[16,17]. In the iron pnictides, however, there is no STM report so far regarding the relationship between the SDW and SC phases[18]. The main obstacle lies in the difficulty of identifying the spectroscopic feature of the SDW phase, namely the SDW gap, in STM studies of the iron pnictides. Recently, the SDW gap was directly observed in the 111-type NaFeAs parent compound due to its excellent cleavage properties[19]. This progress makes the electron doped $NaFe_{1-x}Co_xAs$ a promising candidate for investigating the evolution between magnetism and superconductivity in the underdoped regime.

Figure 1a displays the phase diagram of the $NaFe_{1-x}Co_xAs$ system in the underdoped regime around the SDW and SC phase boundary[20]. The solid symbols in the phase diagram represent the SDW and SC transition temperatures of the three samples studied here. The parent compound ($x = 0$) has a SDW phase transition at $T_{SDW} = 40$ K, and the optimally doped sample ($x = 0.028$) has a bulk SC transition at $T_C = 20$ K. The underdoped sample that lies in the middle ($x = 0.014$) is the most interesting one, which shows a structural transition at $T_S = 32$ K, a SDW transition at $T_{SDW} = 22$ K and a SC transition at $T_C = 16$ K. The three successive transitions are clearly manifested in the specific heat data shown in Fig. 1b and the resistivity results reported previously[20]. The main goal of our STM studies here is to determine whether the SDW-SC coexistence is microscopic or not.

Figure 1c and 1e are the topographic images of the parent and optimally doped NaFe$_{1-x}$Co$_x$As. Figure 1d and 1f display the *dI/dV* curves taken at the base temperature $T = 5$ K along the red lines drawn in the topography. The measured *dI/dV*(*V*,**r**) is proportional to the local electron density of state (LDOS) at sample bias *V* and location **r**. As has been reported before, the parent compound exhibits a large, particle-hole (p-h) asymmetric SDW gap[19]. The optimally doped sample, on the other hand, shows a p-h symmetric SC gap with sharp coherence peaks. In either sample there is only one gap feature in the *dI/dV* curves. This, plus the fact that the *dI/dV* curves are spatially uniform despite the disorders in topography, suggest the existence of a single homogeneous phase for the parent and optimally doped samples. The distinct features of the SDW and SC gaps shown here provide the necessary fingerprints for identifying these two orders in the underdoped sample.

Figure 2a shows a large scale topography of the underdoped $x = 0.014$ sample measured with sample bias $V = 100$ mV and tunneling current $I = 20$ pA. Figure 2b shows the *dI/dV* curves measured at $T = 5$ K along the line drawn in Fig. 2a. There are two striking differences between the spectroscopy in this underdoped sample and that shown above in the parent and optimally doped samples. First, most *dI/dV* spectra clearly exhibit two distinct gap features, a small p-h symmetric gap near the Fermi energy ($E_F$) and a larger p-h asymmetric gap further away from $E_F$. Second, the *dI/dV* curves display pronounced spatial variations. Both the size of the two gaps and amplitude of the gap edge peaks vary considerably from place to place. With increasing $T$ to 10 K (Fig. 2c), the smaller gap becomes much shallower whereas the change of the larger gap is less significant. As $T$ is increased further to 16 K (Fig. 2d), the $T_C$ of this sample, the small gap totally vanishes, indicating it is the SC gap. In contrast, the large gap remains and shows the p-h asymmetric lineshape characteristic of the SDW gap. Figure 2e displays the spatial average of all the *dI/dV* curves taken at the above three temperatures, which demonstrates the evolution from a two-gap structure at $T = 5$ K to a single SDW gap at $T = 16$ K. Thus for 16 K < $T$ < 22 K the FS is partially gapped by the SDW order with a residual FS susceptible for Cooper pairing (there is a large residual DOS near zero bias even in the SDW phase of parent NaFeAs[19]). This strongly resembles the phenomenology of the pseudogap phase in the cuprates.

The SC gap features can be made clearer when the *dI/dV* curves below $T_C$ is divided by that measured at $T_C$ (Ref. 16). Shown in Fig. 3a are such normalized *dI/dV* curves where the data taken at $T = 5$ K are divided by the 16 K ones at the same location. It is evident that the sharp, p-h symmetric SC gap is present in all the spectra. The averaged spectrum in Fig. 3b shows that the two sharp coherence peaks are located at ±5 mV, implying $2\Delta_{SC} = 10$ meV. The normalized 10 K *dI/dV* curves in Fig. 3c show that the SC gap is still present everywhere, although the gap depth is reduced by thermal excitations (Fig. 3d). The existence of the SC gap at every location of the sample demonstrates that superconductivity is a homogeneous rather than a filamentary phenomenon in this underdoped sample. This is supported by the full Meissner shielding fraction in the sample[20]. The direct observation of both the SDW and SC gaps at the same atomic location provides compelling evidence for the microscopic coexistence of the two orders.

Now the microscopic coexistence of the SDW and SC orders is unambiguously demonstrated, the next question is the interplay between them. We answer this question by studying the *dI/dV* curves in further details. Figure 4a zooms into five representative *dI/dV* curves taken at locations with nearly equal spacing along the line in Fig. 2a and marked in Fig. 4c and 4d. First we note that the two distinct gap features is more pronounced for negative bias, where the hump at $V = -17$ mV (marked by blue broken line) is associated with the SDW gap edge and that at $V = -5$ mV (red broken line) is associated with the SC gap edge. Second, when the -17 mV hump is more pronounced, such as in curve 10, 19 and 41, the SC coherence peak at -5 mV is strongly suppressed. On the contrary, in curve 31 and 50 the SC coherence peaks are very sharp, but the SDW hump at -17 mV becomes very weak. Therefore, there exists an obvious anti correlation between the strength of the SDW and SC gap features. Fig. 4b plots the *dI/dV*@-17mV as a function of *dI/dV*@-5mV for all the curves shown in Fig. 2b, which clearly demonstrates the anti correlation between the two quantities. Hence the SDW and SC order competes with each other.

This competition can be visualized directly by utilizing the spectroscopic imaging capability of STM. Shown in Fig. 4c and 4d are the *dI/dV* maps, i.e., the spatial distribution of the electron LDOS, at -17 mV and -5 mV biases for the same field of view as in Fig. 2a. The bright (dark) color represents large (small) LDOS. Except for some sporadic spots with strong impurity states, the bright area of the first map matches very well with the dark area of the second map, and vice versa. The anti correlation between the two *dI/dV* maps presents a direct visualization of the competition between the SDW and SC orders. The competition between magnetism and superconductivity has also been reported by neutron and X-ray diffraction[7], infrared spectroscopy[6] and μSR[9] studies on underdoped 122 compounds.

The direct visualization of the microscopic coexistence and competition between the SDW and SC orders in the underdoped regime puts strong constraints on the theoretical models for the iron based superconductors. The usual reason for not believing the coexistence of a density wave order and superconductivity is because once the FS is gapped by the former the susceptibility toward Cooper pairing is strongly suppressed. For the pnictides it has been demonstrated both theoretically and experimentally that in the SDW ordered state there are residual FSs. However the topology of these FSs is entirely different from that of the non-magnetic state. Since our data have directly demonstrated that Cooper pairing develops when the sample is already partially gapped by the magnetic order, we are driven to the conclusion that Cooper pairing must be developed on top of the residual (magnetically reconstructed) FSs.

Under such condition the usual kinematic argument (for magnetic fluctuation caused Cooper pairing) in the *non-magnetic* state could break down. For example in Ref. 21 it was shown when the SDW ordering moment is large, the residual FSs consists of a hole pocket at the Brillouin zone center and two kidney-shaped electron pockets slighted displaced from it along the SDW ordering wavevector on either sides. For this topology magnetic fluctuations at the usual ordering wavevector cannot even scatter electron between FSs. In this case it is reasonable to believe SDW order will completely suppress SC pairing. On the other hand for weak SDW ordering Ref. 21 shows a residual hole pocket at the Brillouin zone center and two electron pockets at the zone face, whose center momentum is perpendicular to the magnetic ordering wavevector. For this topology it is reasonable to believe magnetic fluctuations at wavevectors perpendicular to the ordering wavevector will cause Cooper pairing. Here magnetic order does not eliminate Cooper pairing and SDW and SC order can coexist in real space. Of course these two orders do not coexist at the same k in momentum space. Therefore, our STM results suggest that SC pairing here is most likely caused by magnetic fluctuations in a SDW phase with a small ordering moment, which has a reconstructed FS with favorable topology. This proposal is supported by the extremely small ordering moment of 0.09 $\mu_B$ detected by neutron scattering in parent NaFeAs[22].

The underlying physics revealed here bears strong resemblance to the pseudogap phenomena in the cuprates. In recent years there are growing evidences that the cuprate pseudogap represents a density wave order that causes a reconstruction of the FS[10-12,16]. The SC phase develops on the residual FSs at low $T$ and the two orders compete with each other[10-12]. Here in underdoped NaFe$_{1-x}$Co$_x$As we demonstrate unambiguously that a SDW gap is present at $T > T_C$. Upon cooling below $T_C$ the SC gap opens on the residual FSs (analogous to the "Fermi arcs" in the cuprates). This important common feature between the pnictides and cuprates suggests that the competition between Cooper pairing and density wave order is generic to strongly correlated electron systems. Microscopically this must mean bare repulsive (local) electron interaction tends to renormalize into effective interactions favoring Cooper pairing and spin/charge density wave order at low energies.

Our results also have indirect implications for the pairing symmetry in the iron pnictides. ARPES data on optimally doped NaFe$_{1-x}$Co$_x$As reveal an isotropic gap function[23], but it cannot distinguish the sign changing $s_\pm$ (Refs 24-27) and the conventional $s_{++}$ symmetry[28]. As shown by Ginzburg-Landau theory analysis and effective model calculations[13,14], the SDW and SC orders almost always mutually exclude each other when the pairing symmetry is $s_{++}$. In contrast, for $s_\pm$ pairing symmetry there is a wide parameter range in which the two orders can microscopically coexist. Thus we take our STM results as an indirect support of the $s_\pm$ pairing symmetry. Of course only a phase sensitive experiment, such as the magnetic field-dependent quasiparticle interference[29], can prove it definitively.

## Method Summary

**Sample growth.**

High quality NaFe$_{1-x}$Co$_x$As single crystals are grown by the self flux technique as described elsewhere[20]. NaAs precursor is firstly synthesized at 200 °C for 10 h, and then powders of NaAs, Fe and Co are mixed together according to the ratio NaAs : Fe : Co = 4 : 1-$x$ : $x$. The mixture is placed in an alumina crucible and then sealed into an iron crucible. The samples are put in a tube furnace with inert atmosphere and melt at 950 °C for 10 h before slowly cooled down to 600 °C at a rate of 3 °C/h. The actual chemical composition of the single crystals is determined by X-ray energy dispersive spectroscopy (EDS).

**STM measurements.**

The STM experiments are performed with a low temperature ultrahigh vacuum (UHV) STM system. The NaFe$_{1-x}$Co$_x$As crystal is cleaved *in situ* at $T = 77$ K (the cleaving stage is cooled by liquid nitrogen) and then transferred immediately into the STM sample stage. An electrochemically etched polycrystalline tungsten tip is used in the experiments. The STM topography is taken in the constant current mode, and the *dI/dV* spectra are collected using a standard lock-in technique with modulation frequency $f = 423$ Hz.

**Acknowledgments** This work was supported by the National Natural Science Foundation and MOST of China (grant No. 2009CB929400, 2010CB923003, 2011CBA00101, and 2012CB922002). D.H.L. was supported by DOE grant number DE-AC02-05CH11231.

**Author Information** Correspondence and requests for materials should be addressed to Y.W. (yayuwang@tsinghua.edu.cn).

Figure Captions:

**Figure 1 | Phase diagram of NaFe$_{1-x}$Co$_x$As and STM results on parent and optimally doped samples. a,** Schematic electronic phase diagram of NaFe$_{1-x}$Co$_x$As in the underdoped regime. The solid symbols mark the SDW and SC transition temperatures of parent ($x = 0$), underdoped ($x = 0.014$), and optimally doped ($x = 0.028$) samples. **b,** Temperature dependence of the specific heat and its derivative (inset) of the underdoped sample show the structural, SDW and SC phase transitions. **c,** Constant current image of the cleaved surface of parent NaFeAs. **d,** The *dI/dV* spectroscopy taken on the parent compound along the red line in **c**. There is a single p-h asymmetric SDW gap that is uniformly distributed in space. **e,** Topography of the optimally doped sample, which shows more disorders due to Co doping. **f,** The *dI/dV* spectroscopy taken on the optimally doped sample along the red line in **e** shows a single p-h symmetric SC gap that is also quite homogeneous.

**Figure 2 | The *dI/dV* spectra of the underdoped NaFe$_{1-x}$Co$_x$As. a,** A large area topography of the underdoped $x = 0.014$ sample taken with $V = 100$mV and $I = 20$ pA. **b,** The *dI/dV* spectra taken at $T = 5$ K along the red line drawn in **a**. The spectra are much more complex than that in the parent and optimally doped samples. They present strong spatial variations, and in most spectra there are clearly two gap features at the same atomic location. **c, d,** The same linecut spectra taken at $T = 10$ K and 16 K, respectively. The small gap features become significantly weaker at 10 K and vanish completely at 16 K, indicating that it is the SC gap. The large gap persists

to 16 K and shows the p-h asymmetric lineshape characteristic of the SDW gap. **e,** The spatial average of all the curves taken at different *T*s, showing the evolution from the two-gap structure at 5 K to a single SDW gap at 16 K.

**Figure 3 | The SC gap extracted by the normalization. a,** The *dI/dV* curves taken at 5 K are divided by that taken at the same location at 16 K. The SC gap feature is present in every location of the sample, indicating it is a bulk phase. **b,** The spatially averaged spectrum shows that the coherence peaks are located at $V = \pm 5$ mV. **c,** Normalized spectra and **d,** its spatial average of the 10 K *dI/dV* curves. The SC gap features are still present everywhere although the gap is significantly filled by thermal excitations.

**Figure 4 | Anti-correlation between the SDW and SC features. a,** Close-up of 5 representative *dI/dV* curves taken at the locations marked in **c** and **d**. The two-gap features are more pronounced in the negative bias side where the SDW gap has a gap edge peak at -17 mV and SC gap has a coherence peak at -5 mV. **b,** The *dI/dV* value at -17 mV as a function of that at -5 mV for all the curves in Fig. 2b. There is an obvious anti-correlation between the two quantities indicated by the red broken line, which can also be seen directly in the 5 curves in **a**. **c** and **d**, The *dI/dV* maps measured at -5 mV and -17 mV on the same field of view as Fig. 2a. The brightness of the two maps, which corresponds to the strength of SC and SDW phases, shows clear anti-correlation.

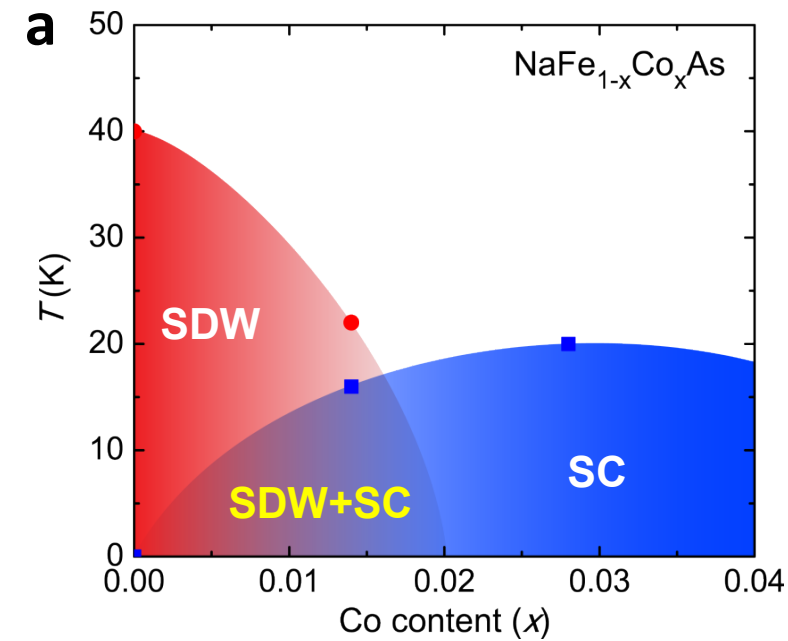
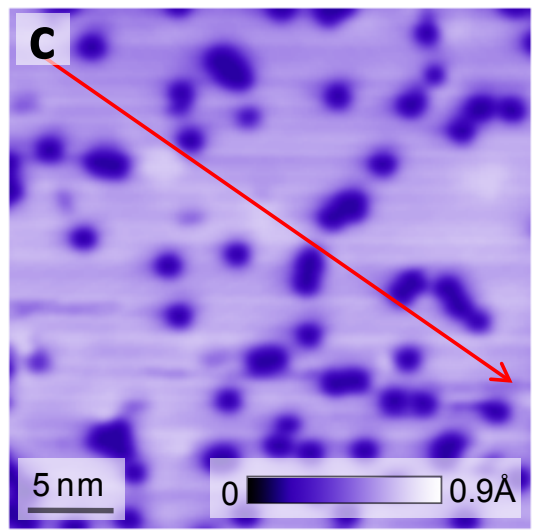
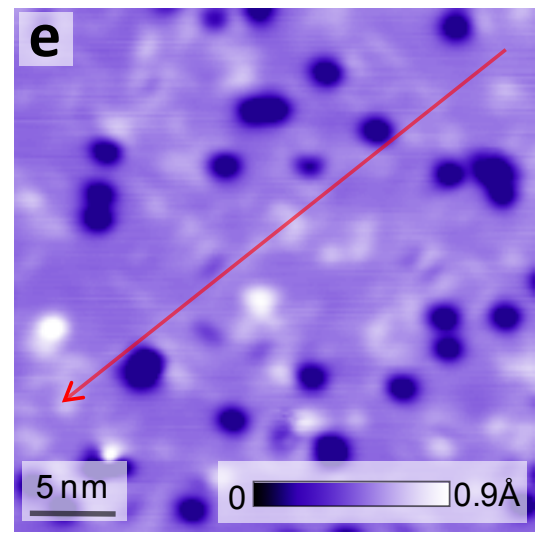
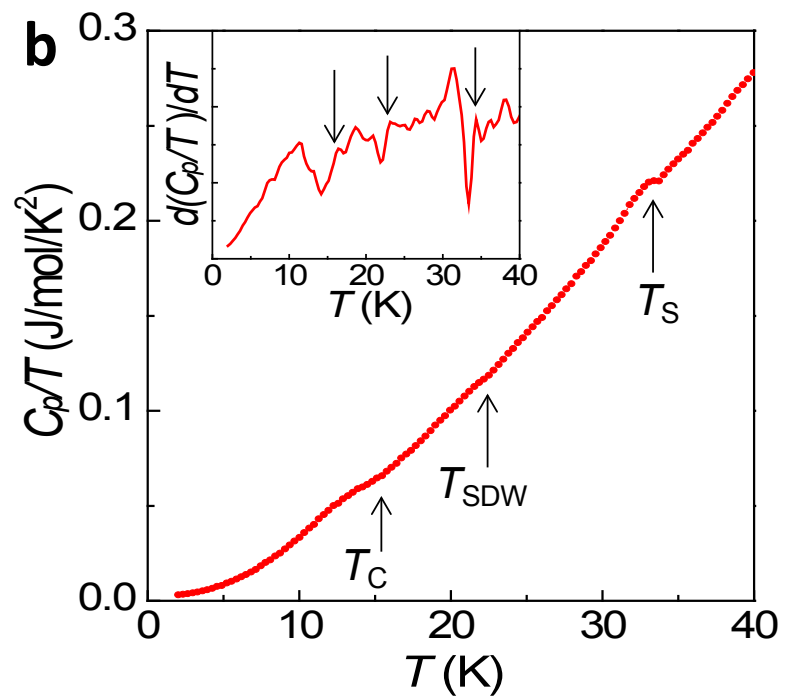
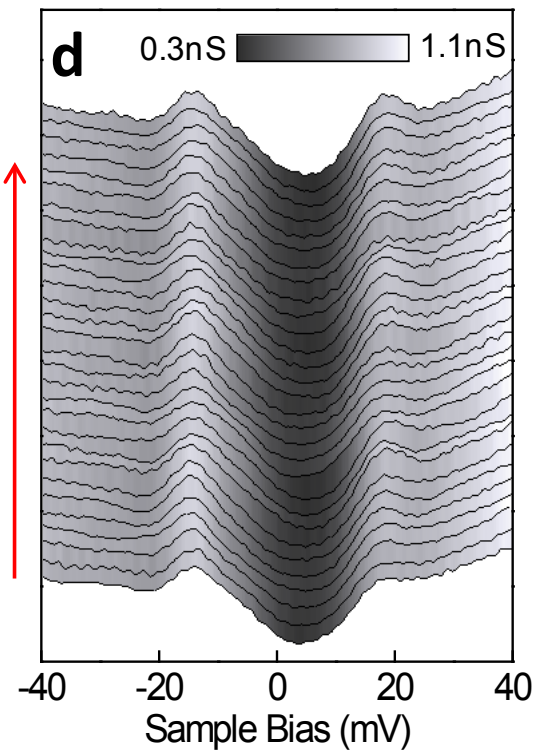
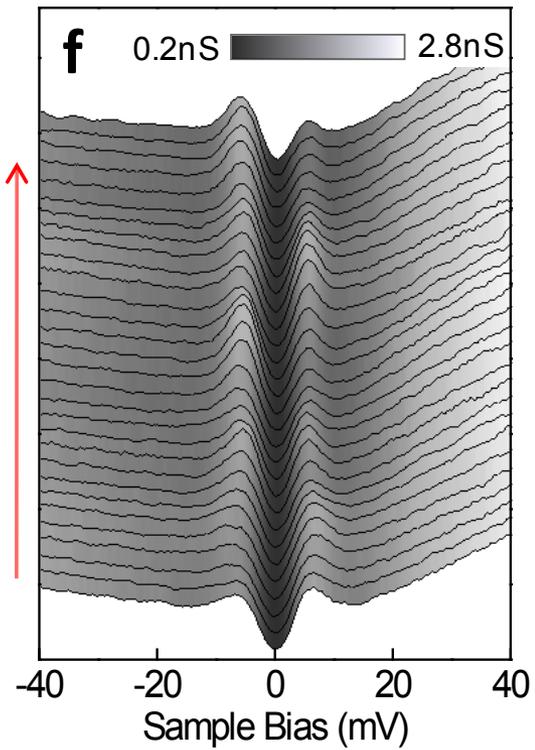

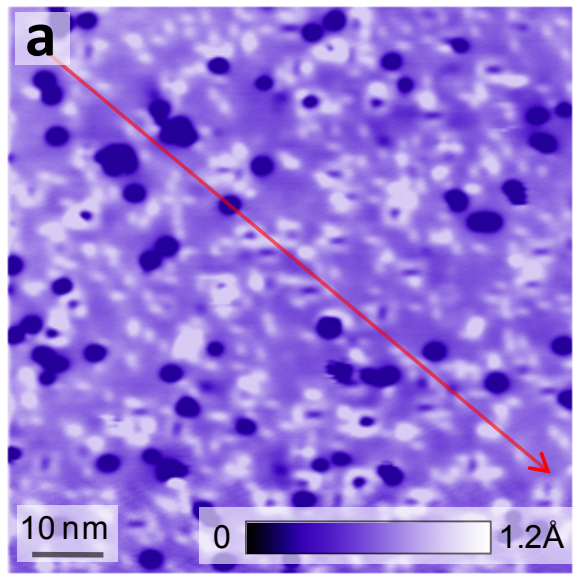
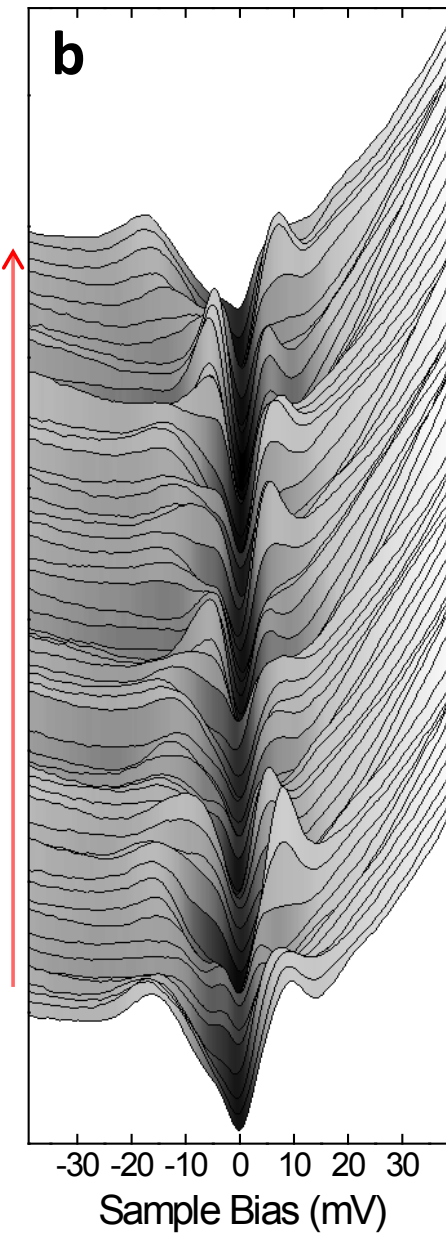
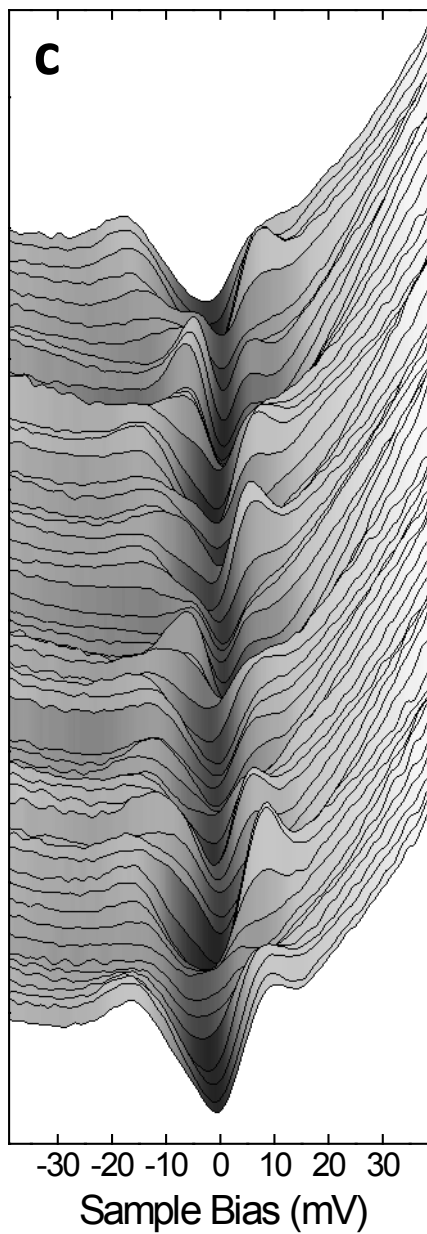
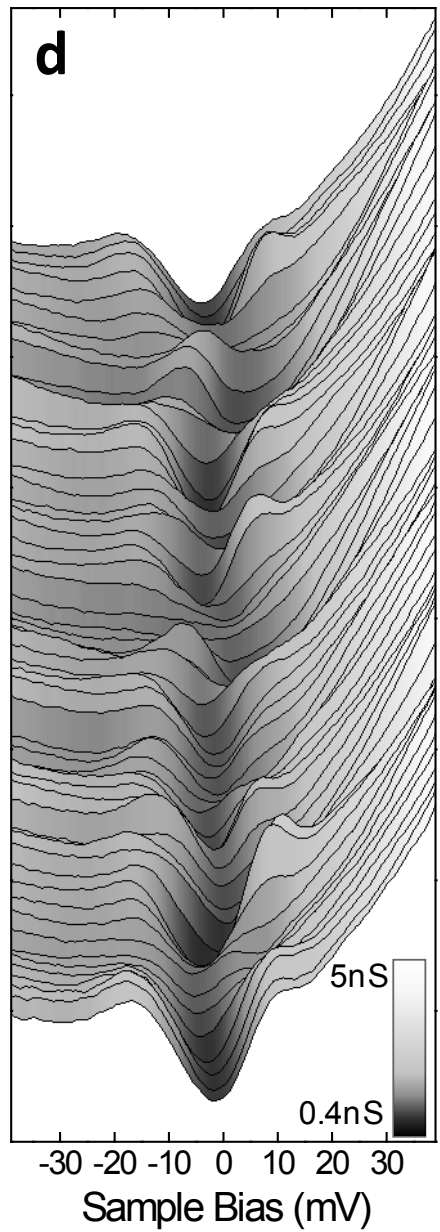
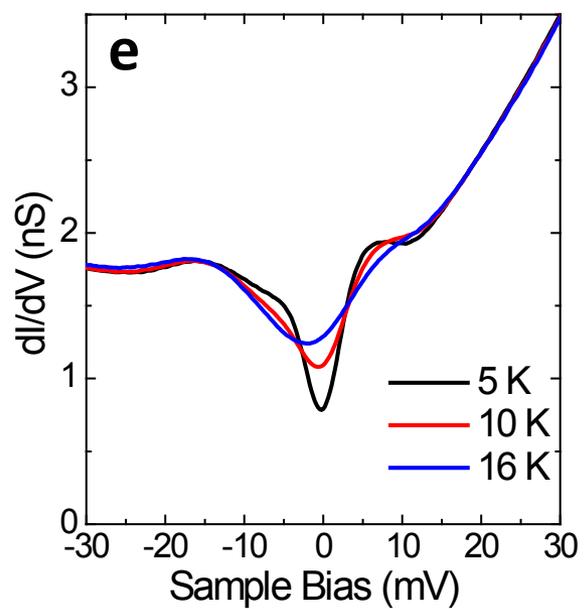

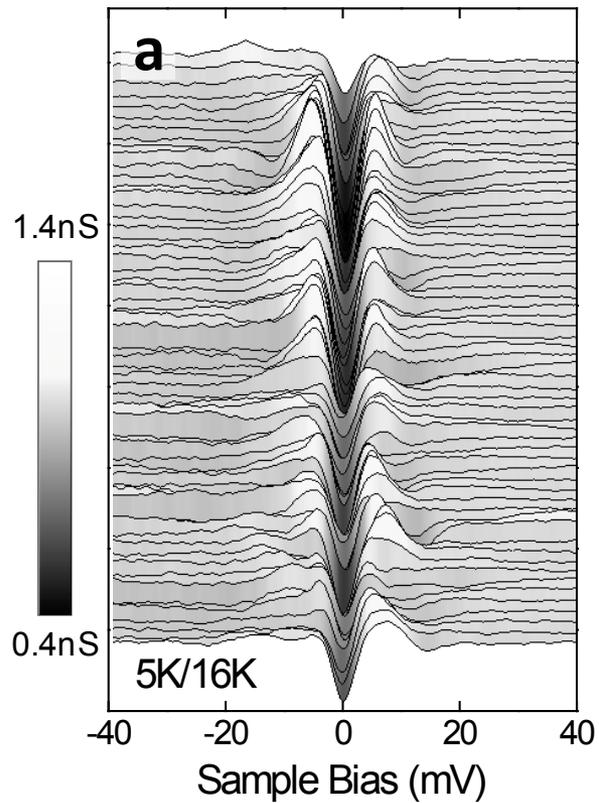
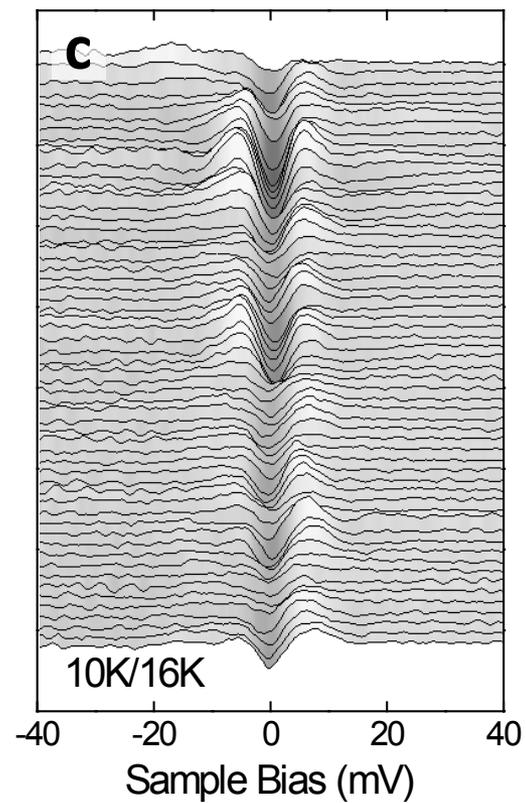
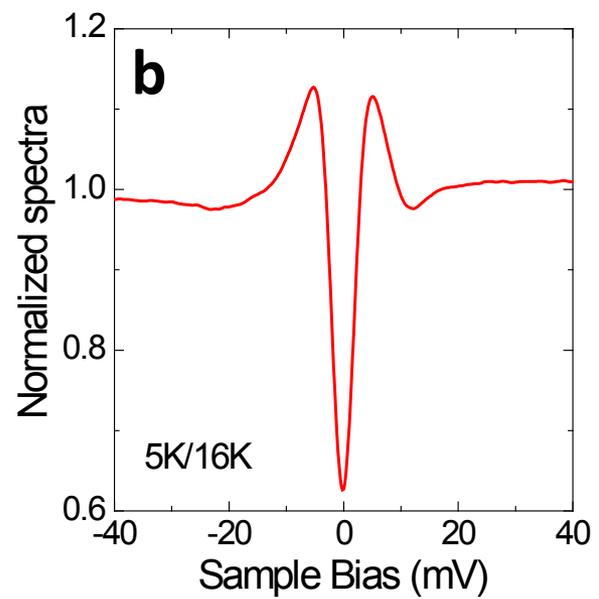
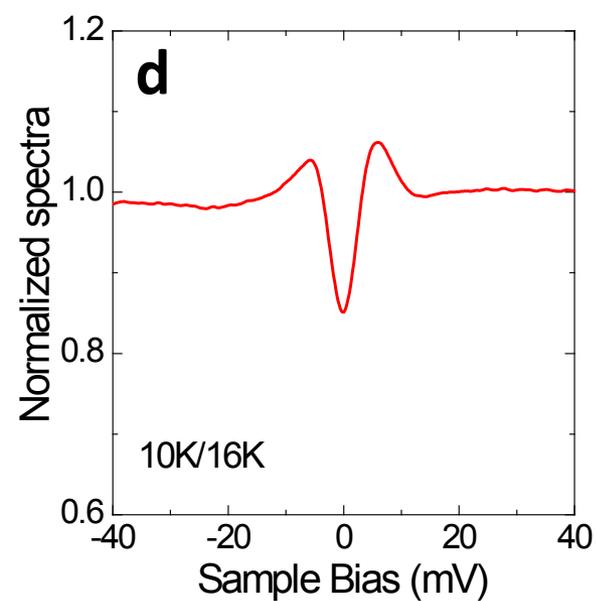

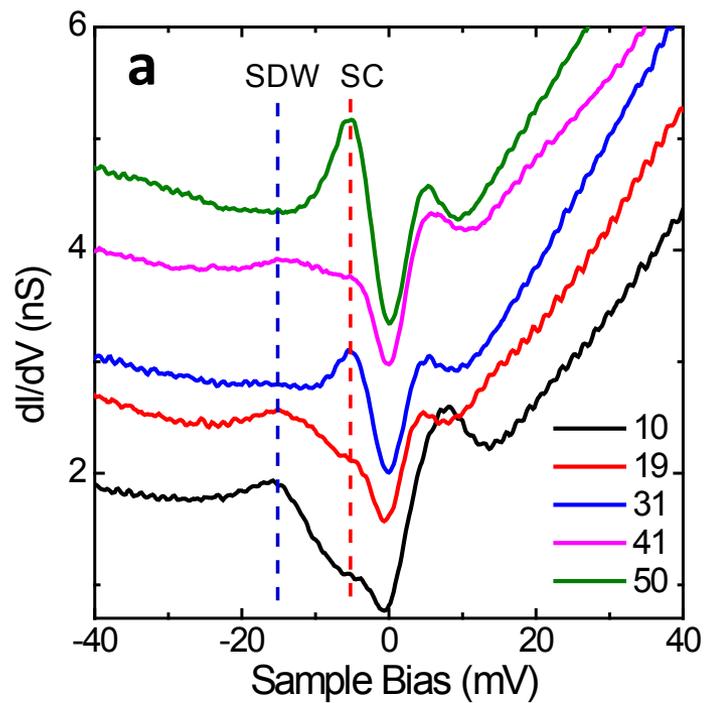
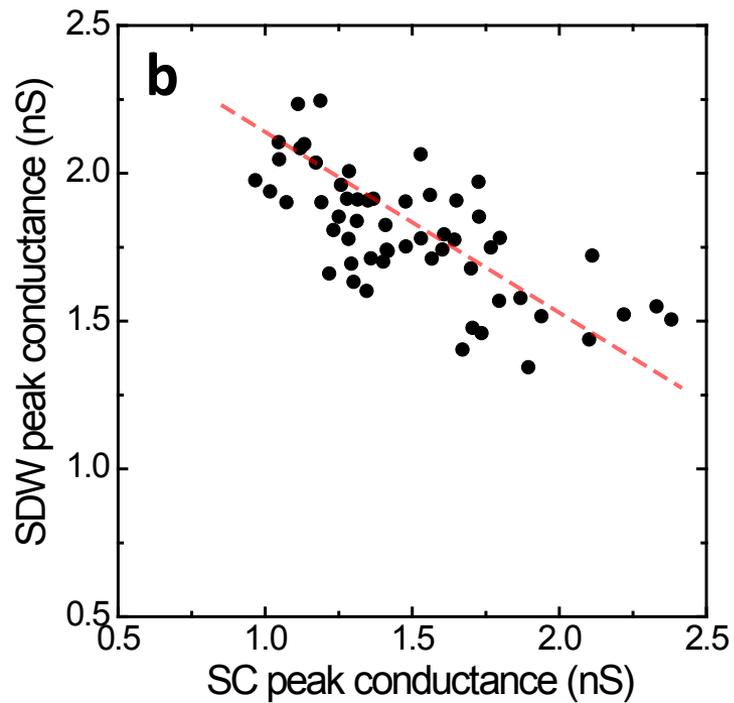
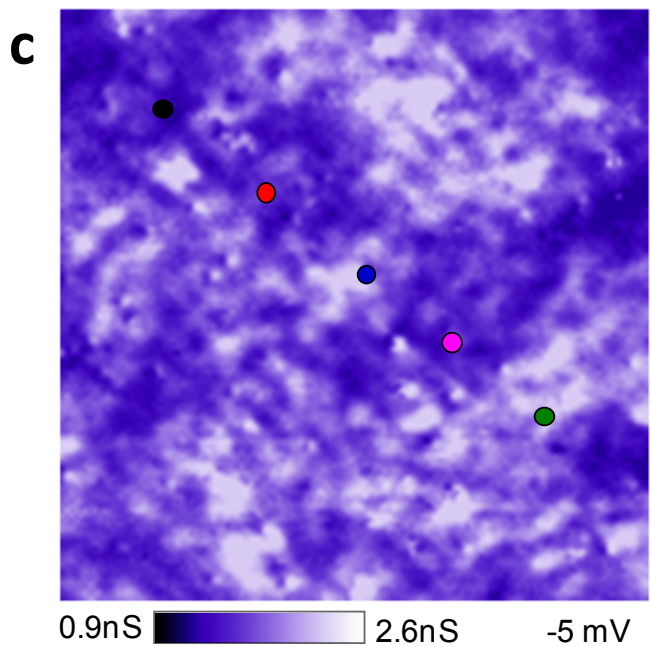
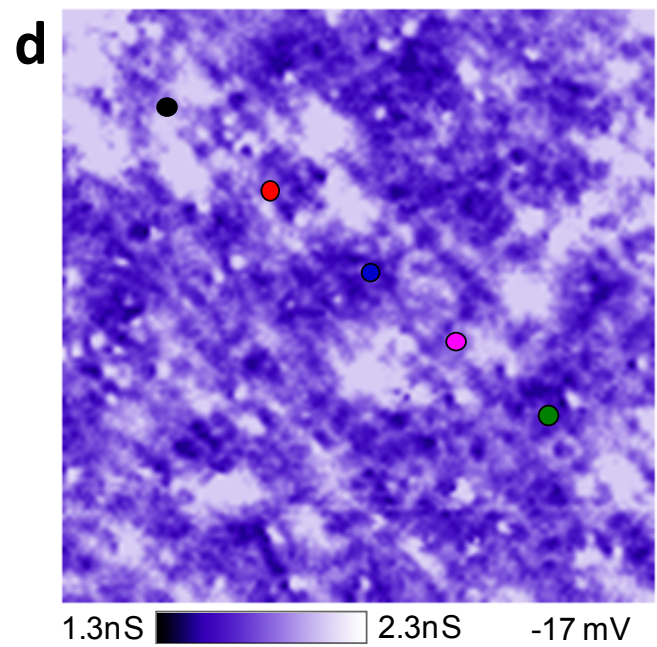